\documentclass[conference]{IEEEtran}

\newcommand{\ignore}[1]{}
\usepackage{fancyhdr}
\usepackage[normalem]{ulem}
\usepackage[hyphens]{url}
\usepackage{balance}

\newcommand{\code}[1]{\texttt{#1}}

\usepackage{times}
\usepackage{microtype}
\usepackage{mathtools}

\usepackage{listings}
\usepackage{color}
\usepackage{algorithm, algorithmic}
\usepackage{blindtext}
\usepackage{cite}
\usepackage{color}   
\usepackage{amssymb,amsmath}
\usepackage{comment}
\usepackage{enumerate}
\usepackage{framed}
\usepackage{array}      
\usepackage{url}		
\usepackage{balance}
\usepackage{xspace}
\usepackage{url}         
\usepackage{multirow}
\usepackage{listings}
\usepackage{booktabs}
\usepackage{amsmath}
\usepackage{datetime}
\usepackage{subcaption}
\usepackage{amsmath}
\usepackage{graphicx}
\usepackage{siunitx}
\usepackage{textcomp}

\newcommand{\bluecomment}[1]{\textcolor{blue}{\{#1\}}}

\fancypagestyle{firstpage}{
  \fancyhf{}
\setlength{\headheight}{50pt}

\fancyhead[C]{\normalsize{DATE 2018 Submission
       -- Confidential Draft -- Do NOT Distribute!!} \\
      \textbf{Generated on: \today\ \currenttime} 
  }
  \pagenumbering{arabic}
}

\graphicspath{{figures/}{pics/}}
\DeclareGraphicsExtensions{.pdf,.jpeg,.png}

\title{Correlation Manipulating Circuits for \\Stochastic Computing}

\author{Vincent T. Lee, Armin Alaghi, Luis Ceze \\
  University of Washington \\
  \{vlee2, armin, luisceze\}@cs.washington.edu
  \vspace{-2mm}
}

\begin{document}
\sloppy
\maketitle
\pagestyle{plain}
\bstctlcite{IEEEexample:BSTcontrol}
\pagenumbering{gobble}


\begin{abstract}

  Stochastic computing (SC) is an emerging computing technique that promises high density, low power, and error tolerant solutions.
  In SC, values are encoded as unary bitstreams and SC arithmetic circuits operate on one or more bitstreams.
  In many cases, the input bitstreams must be correlated or uncorrelated for SC arithmetic to produce accurate results.
  As a result, a key challenge for designing SC accelerators is manipulating the impact of correlation across SC operations.
  This paper presents and evaluates a set of novel correlation manipulating circuits to manage correlation in SC computation: a synchronizer, desynchronizer, and decorrelator.
  We then use these circuits to propose improved SC maximum, minimum, and saturating adder designs.
  Compared to existing correlation manipulation techniques, our circuits are more accurate and up to 3$\times$ more energy efficient.
  In the context of an image processing pipeline, these circuits can reduce the total energy consumption by up to 24\%.

\end{abstract}

\vspace{2mm}

\section{Introduction}
\label{sec:introduction}

\noindent The search for high density and low power solutions has accelerated interest towards a wide variety of alternative computing techniques and technologies.
Stochastic computing (SC) is one such emerging computing technique that has recently enjoyed renewed interest because of its compact size, low power, and improved error tolerance.
Unlike binary-encoded (BE) computation, SC uses unary-encoded bitstreams (time series of 1s and 0s) to encode values~\cite{gaines69}.

In SC, the value of a bitstream is encoded by the number of constituent 1s and 0s, and the precision of the bitstream is governed by the bitstream length.
For instance, the bitstream $X = 01000100$ encodes the value $p_X = 0.25$ since there are two 1s and the bitstream length is 8.
This encoding allows for area efficient and low power implementations of arithmetic primitives such as multiplication.
For instance, given two bitstreams $X = 01010101$ $(p_X=0.5)$, $Y = 11111100$ $(p_Y=0.75)$, we can compute the product $Z = 01010100$ $(p_Z=0.375)$ by calculating the bitwise AND of $X$ and $Y$.
Notice that the multiplication is only accurate because the inputs $X$ and $Y$ are generated independently and hence \textit{uncorrelated}.
Also notice that what SC circuits gain in power and area, they lose in performance: run time in SC is proportional to the length of the bitstream.

Unlike BE computation, many operations in SC do not yield exact results.
There are two primary sources of errors in SC: errors due to quantization, and errors due to insufficient or excessive correlation between input operand bitstreams.
In SC, the input and output precision of arithmetic operations is the same; as a result, any arithmetic operation which requires higher output precision than input precision (e.g. addition) results in quantization errors in SC.

\begin{table}[t]
  \caption{SC functions implemented by a two-input AND gate. Different correlation results in different functions.}
  \label{tab:correlation-example}
  \centering
  \begin{tabular}{@{}m{0.17\linewidth}m{0.12\linewidth}m{0.12\linewidth}m{0.12\linewidth}m{0.28\linewidth}@{}} \toprule
    \centering Inputs & \centering $X$ & \centering $Y$ & \centering $X \& Y$ & \centering Function \tabularnewline \midrule
    \centering \begin{tabular}{@{}c@{}}Positively\\Correlated\end{tabular} &
      \begin{tabular}{@{}c@{}}10101010\\(0.5)\end{tabular} &
      \begin{tabular}{@{}c@{}}10111011\\(0.75)\end{tabular} &
      \begin{tabular}{@{}c@{}}10101010\\(0.5)\end{tabular} & \centering $\code{min}(p_X, p_Y)$ \tabularnewline \midrule
    \centering \begin{tabular}{@{}c@{}}Negatively\\Correlated\end{tabular} &
      \begin{tabular}{@{}c@{}}10101010\\(0.5)\end{tabular} &
      \begin{tabular}{@{}c@{}}11011101\\(0.75)\end{tabular} &
      \begin{tabular}{@{}c@{}}10001000\\(0.25)\end{tabular} & \centering $\code{max}(0,p_X+p_Y-1)$ \tabularnewline \midrule
      \centering \begin{tabular}{@{}c@{}}Uncorrelated\\Inputs\end{tabular} &
        \begin{tabular}{@{}c@{}}10101010\\(0.5)\end{tabular} &
        \begin{tabular}{@{}c@{}}11111100\\(0.75)\end{tabular} &
        \begin{tabular}{@{}c@{}}10101000\\(0.375)\end{tabular} & \centering $p_X\times p_Y$ \tabularnewline \bottomrule
  \end{tabular}
\end{table}

In contrast, errors due to correlation arise because of over-correlated or insufficiently correlated SC input operands; many SC circuits require correlated or uncorrelated inputs to operate correctly.
Table~\ref{tab:correlation-example} shows the set of implementable functions using a 2-input AND gate where input operands have the same value but different correlation levels.
Notice that uncorrelated input bitstreams correctly calculate multiplication while the two other cases with positively/negatively correlated input bitstreams realize different functions.
Manipulating and managing the impact of correlation is the focus of this paper.

This paper presents a set of new SC correlation manipulating circuits: a synchronizer and desynchronizer for increasing positive and negative correlation respectively, and a decorrelator for reducing correlation.
Existing correlation manipulation techniques rely on correlated or uncorrelated random number generators when bitstreams are generated~\cite{sc-correlation}; this can be prohibitively expensive since converting bitstreams to and from the BE domain requires significant power and area overheads.
Instead, the correlation manipulating circuits proposed in this work can be inserted at appropriate points in the computation and are more energy efficient than converting to and from the BE domain (discussed later).
We also propose improved SC maximum and minimum circuits using our synchronizer, and an improved SC saturating adder using our decorrelator; we also demonstrate the utility of our correlation manipulating circuits in the context of an image processing pipeline.

Our contributions are as follows:
(1) A novel synchronizer and desynchronizer circuit for \textit{increasing} positive and negative correlation respectively between two SC bitstreams.
(2) A novel decorrelator circuit for \textit{reducing} correlation between two SC bitstreams.
(3) An improved SC maximum and minimum design using synchronizers, and an improved saturating adder using desynchronizers.




The paper is organized as follows:
Section~\ref{sec:background} provides background on SC.
Section~\ref{sec:design} presents our new correlation manipulating circuits and improved SC circuits.
Section~\ref{sec:evaluation} evaluates correlation manipulating circuits in the context of an image processing pipeline.
Section~\ref{sec:related-work} presents related work.

\section{Background}
\label{sec:background}

\noindent This section provides background on SC and discusses the role of correlation.

\subsection{Stochastic Computing Basics}

\noindent Stochastic computing (SC) is an alternative computing technique introduced in the 1960s~\cite{gaines69} and uses unary bitstreams (time series of 1s and 0s) to represent values.
The value of a bitstream is encoded by the number of 1s and 0s, and the precision is determined by the bitstream length.
SC bitstreams are often referred to as \textit{stochastic numbers} (SNs) and either use unipolar or bipolar encodings.

In unipolar-encoded SNs, 1s have a weight of +1 and 0s have a weight of 0; the encoded value is the sum of each position in the SN divided by the SN length $N$, and is limited to the range [0, 1].
For instance, the SN $X = 01100001$ has the value $p_X = 3/8$ since there are three 1s.
Alternatively, bipolar-encoded SNs weight 1s as +1 and 0s as $-$1 allowing them to encode both negative and positive values in the range [$-$1, +1].
For example, the same SN $X = 01100001$ has the value $p_X = -1/4$ under bipolar encodings.

\begin{figure}[b]
  \centering
  \includegraphics[width=\linewidth]{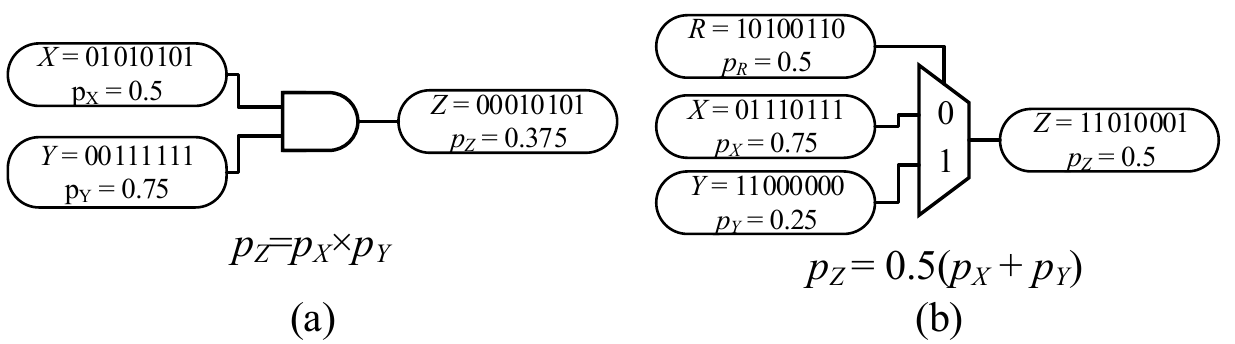}
  \caption{Example operation of SC (a) multiplication, and (b) addition.}
  \label{fig:sc_example}
\end{figure}

The key strength of SC circuits is that they are extremely dense, low power, and error tolerant.
For instance, an SC multiplication can be implemented by a single AND gate (Fig.~\ref{fig:sc_example}a) since, given two SNs $X$ and $Y$ with SC values $p_X$ and $p_Y$ respectively, the bitstream $Z = X \& Y$ has value $p_Z = p_Xp_Y$.
Notice that the SNs $X$ and $Y$ are assumed to be uncorrelated, otherwise the fidelity of the SC multiplication breaks down.
SC addition is implemented using a multiplexor with input SN operands as the data inputs, and an uncorrelated auxiliary SN with value $p_R=0.5$ as the select signal.
The SC adder effectively samples each input SN with equal probability resulting in a scaled sum $p_Z = 0.5(p_X+p_Y)$.
An example of SC addition is shown in Fig.~\ref{fig:sc_example}b.

Notice that in SC, all bits have equal weight which means the equivalent precision of a SN with length $N$ is approximately $log_2(N)$ since it can represent the values $\{0/N, 1/N, 2/N, ..., (N-1)/N, N/N\}$.
Consequently, the input and output SNs for all arithmetic operations must have the \textit{same} precision.
This highlights one source of error in SC: quantization errors due to precision reduction.
For instance, consider SC addition which has a scale factor of 0.5 in the resulting sum; this forces the output SN precision to be the same as the input SN precision, and as a byproduct drops the least significant bit of the true sum.
However, fatal levels of precision reduction can often be avoided by increasing the bitstream length or using higher precision conversion circuits such as accumulative parallel counters (APC)~\cite{sc-apc}.

To generate SNs, SC uses digital-to-stochastic (D/S) converters (Fig.~\ref{fig:sc_primitives}g) which takes the BE integer value $x \in [0, N]$, and a random integer value $r \in [0, N]$ produced by a random number generator (RNG).
The RNG value $r$ is compared against $x$ to produce the desired SN with value $p_X = x / N$.
The choice of RNG is vitally important since correlated or uncorrelated RNGs can be used to generate correlated or uncorrelated SNs respectively.
To convert a SN back to the BE domain, we use a stochastic-to-digital (S/D) converter (Fig.~\ref{fig:sc_primitives}f) which is realized by a counter that sums up each bit in the SN.
S/D converters, D/S converters, and RNGs are overheads unique to operating in the SC domain, and individually they are much larger and consume more power than SC arithmetic circuits.
However, these overheads can be amortized over many SC operations by judiciously exploiting application data reuse and composing multiple operations together.

\subsection{Correlation Sensitivity}

{
  \setlength{\belowcaptionskip}{-15pt}
\begin{figure*}[ht]
  \centering
  \begin{tabular}{@{}m{0.10\linewidth}m{0.11\linewidth}m{0.11\linewidth}
      m{0.11\linewidth}m{0.10\linewidth}m{0.10\linewidth}
      m{0.11\linewidth}m{0.1\linewidth}@{}} \toprule
    \centering Operation & \centering (a) Add &
      \centering (b) Saturating Add        &
      \centering (c) Subtract &
      \centering (d) Multiply &
      \centering (e) Divide &
      \centering (f) S/D Converter &
      \centering (g) D/S Converter \tabularnewline \midrule

      \centering Circuit     &
      \centering \includegraphics[width=\linewidth]{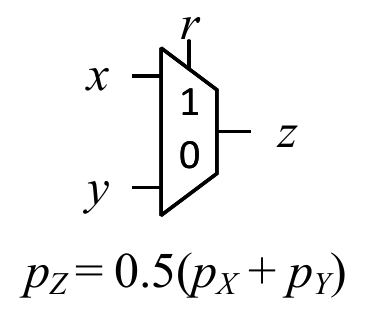} &
      \centering \includegraphics[width=\linewidth]{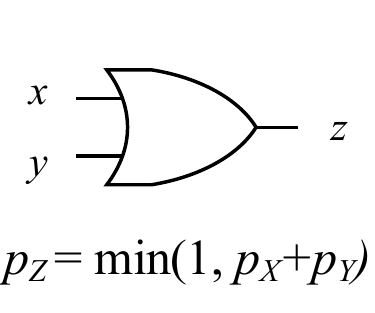} &
      \centering \includegraphics[width=\linewidth]{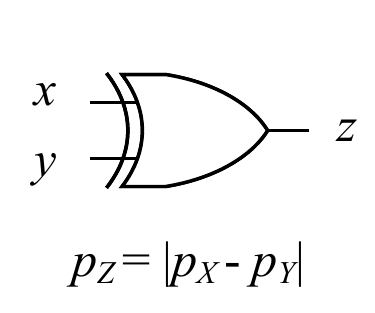} &
      \centering \includegraphics[width=\linewidth]{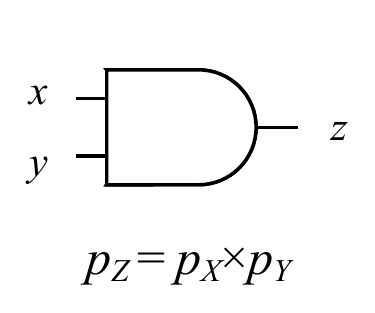} &
      \centering \includegraphics[width=\linewidth]{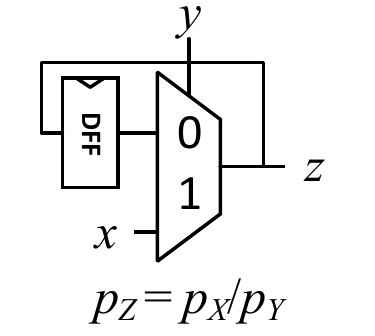}  &
      \centering \includegraphics[width=\linewidth]{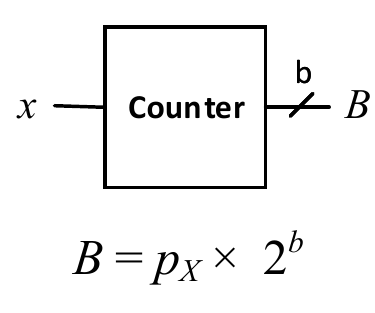}  &
      \centering \includegraphics[width=\linewidth]{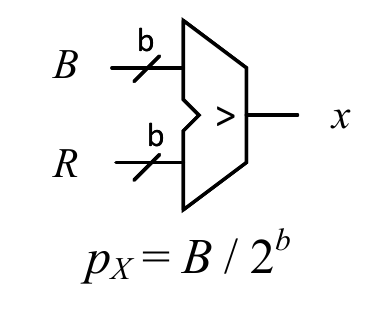}
      \tabularnewline \hline



      \centering Reference & \centering ~\cite{gaines67} & \centering ~\cite{sc-correlation} & \centering ~\cite{alaghi14-date} & \centering ~\cite{gaines67} & \centering ~\cite{sc-divider} \tabularnewline \hline

      \centering \begin{tabular}{@{}c@{}} Operand\\Correlation\end{tabular} &
        \centering Uncorrelated with $r$ &
        \centering \begin{tabular}{@{}c@{}}Negative\end{tabular} &
        \centering \begin{tabular}{@{}c@{}}Positive\end{tabular} &
        \centering Uncorrelated &
        \centering \begin{tabular}{@{}c@{}}Positive\end{tabular} &
        \centering Agnostic & \centering N/A \tabularnewline \bottomrule
  \end{tabular}
  \caption{Correlation sensitive SC operations and converter circuits: (a) scaled add, (b) saturating add, (c) subtract, (d) multiply, (e) divide, (f) S/D converter, and (g) D/S Converter.}
  \label{fig:sc_primitives}
\end{figure*}
}

\noindent Managing and manipulating correlation between SNs in SC computation is critical for producing accurate results.
Fig.~\ref{fig:sc_primitives} shows the set of known arithmetic correlation sensitive circuits; note how some SC operations are more accurate when their input SNs are correlated.
There are three ways to mitigate the detrimental impact of correlation in SC: (1) use correlated or uncorrelated RNGs to generate SNs, (2) use correlation agnostic circuits, or (3) use correlation manipulating circuits to reduce or enhance correlation between SNs.
In SC, the correlation between two SNs $X$ and $Y$ is quantified using the SC correlation (SCC) defined in~\cite{sc-correlation} as:
\vspace{-1mm}
\[
  \label{eq:scc}
  \code{SCC}(X, Y) = \left\{\begin{array}{ll}
  \frac{ad-bc}{N \times \code{min}(a+b, a+c) - (a+b)(a+c)} & ad > bc \\
  \frac{ad-bc}{(a+b)(a+c) - N \times \code{max}(a-d,0)} & \code{else} \\
  \end{array} \right.
\]

\noindent In this definition, $a$ is the number of positions where $X$ and $Y$ are both 1, $b$ is the number of positions where $X$ is 1 and $Y$ is zero, $c$ is the number of positions where $X$ is 0 and $Y$ is 1, and $d$ is the number of positions where $X$ and $Y$ are both 0.
A SCC = 0 means the SNs $X$ and $Y$ are perfectly uncorrelated, while a SCC = +1 and SCC = $-$1 indicates maximal positive and negative correlation respectively. 

As mentioned before, the choice of RNG for generating SNs is important since two SNs generated by uncorrelated RNGs will also be uncorrelated.
Similarly, SNs produced by the same RNG will be positively correlated.
Previous SC work uses linear feedback shift registers (LFSRs) which are simple and compact; however, not all LFSR combinations generate completely uncorrelated SNs.
As a result, it is often necessary to use rotated LFSR outputs or different LFSR seeds to minimize correlation.
More recent work has shown that it is sufficient to use low-discrepancy sequences such as Van der Corput (VDC), Halton~\cite{low-discrepancy-sequences}, or Sobol sequences~\cite{liu17-date}.
Unfortunately, it is impractical to use distinct uncorrelated RNGs to generate all SNs since RNGs are significantly larger and higher power than individual SC operations; as a result, most SC computation amortizes the cost of RNGs by generating many SNs from each RNG.
A key limitation of this technique is that correlation between SNs can only be induced during D/S conversion and can not be used to affect the correlation of intermediate SNs.

There exist a handful of correlation agnostic SC operations which always compute an accurate result regardless of the correlation between input SNs.
Unfortunately, correlation agnostic variants do not exist for all SC operations.
The known set of correlation agnostic circuits are also larger and consume more power than then their equivalent correlation sensitive counterparts.
For instance, in our experiments we find that the correlation agnostic adder introduced in~\cite{lee17-date} is 5.6$\times$ larger and requires 10.7$\times$ more power than the SC adder in Fig.~\ref{fig:sc_primitives}a.


Correlation between SNs is also introduced through SC operations across successive computations.
Unfortunately, the quantitative impact of how each SC arithmetic operation changes the SN correlation with respect to other SNs is not well-understood.
As a result, it is sometimes difficult or impractical to completely guarantee correlated or uncorrelated input SNs across many operations.
To correct correlation levels at intermediary stages of SC computation, we use correlation manipulating circuits.

One way to mitigate the impact of computation-induced correlation is to convert all SNs back to the BE domain using S/D converters and re-encode them to SNs using D/S converters~\cite{ting16}.
This technique is known as \textit{regeneration} since it regenerates SNs to reset any correlation that may have existed or did not exist between SNs.
However, regeneration is expensive to execute since S/D converters and D/S converters consume one to two orders of magnitude more power and area than SC arithmetic circuits.
A smaller and lower power alternative to regeneration is inserting isolators~\cite{ting16}; however, isolators do not modify the order of bits in a SN and can have limited impact on SCC (shown later).
In this work, we introduce new SC correlation manipulating circuits to make such correlation manipulation overheads smaller, more effective, and more energy efficient.

\section{Design and Implementation}
\label{sec:design}

\noindent We now introduce the design of our new correlation manipulating circuits and improved SC operators.

\subsection{Synchronizer and Desynchronizer}

\begin{figure*}[!ht]
  \centering
  \begin{subfigure}{0.5\linewidth}
    \centering
    \includegraphics[width=\linewidth]{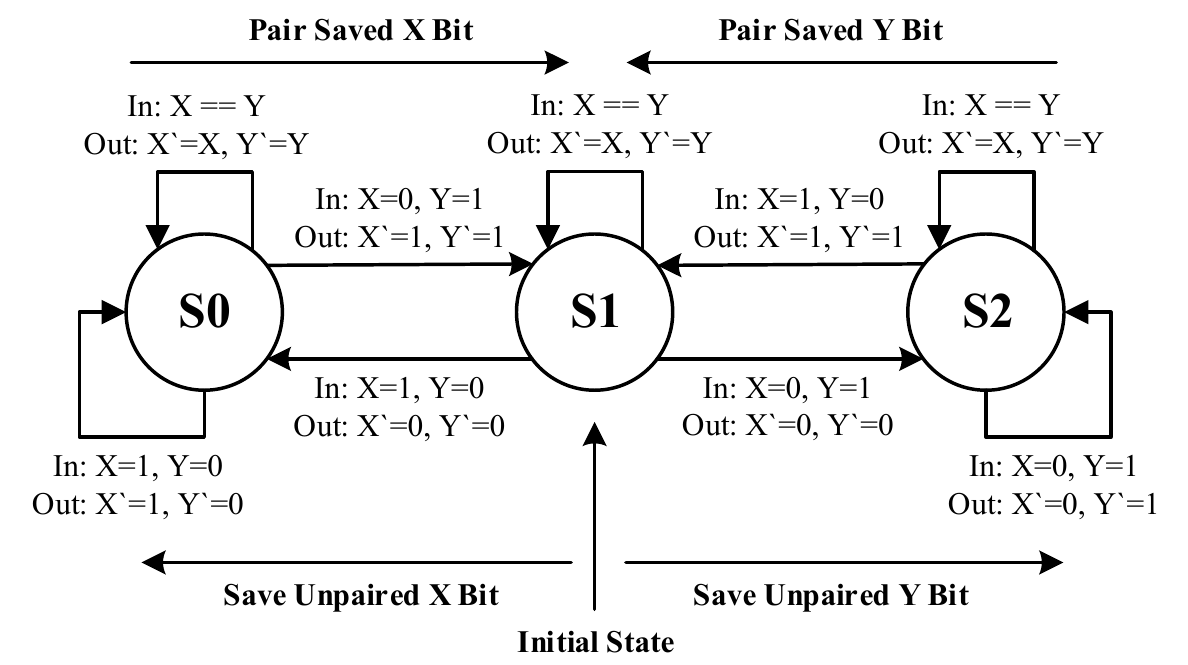}
    \caption{Synchronizer instance with save depth $D = 1$.}
    \label{fig:synchronizer}
  \end{subfigure}%
  \begin{subfigure}{0.5\linewidth}
    \centering
    \includegraphics[width=\linewidth]{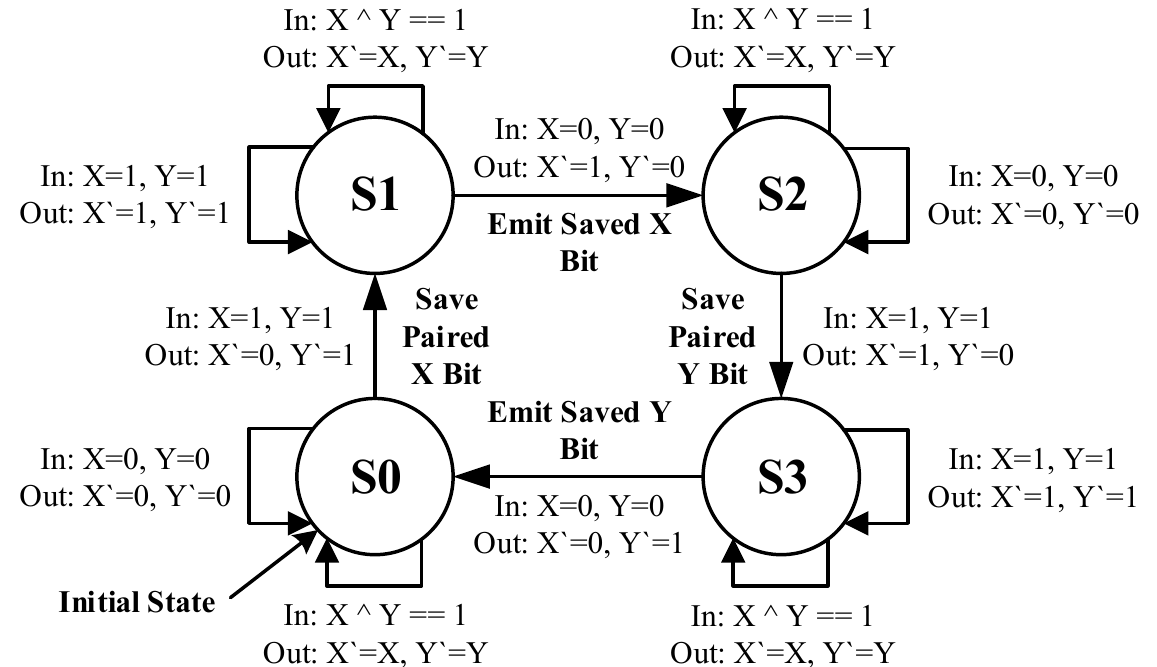}
    \caption{Desynchronizer instance with save depth $D = 1$.}
    \label{fig:desynchronizer}
  \end{subfigure}
  \captionsetup{skip=-0pt}
  \setlength{\belowcaptionskip}{-10pt}
  \caption{Correlation inducing SC designs: (a) synchronizer, and (b) desynchronizer.}
\end{figure*}

\noindent Our novel SC synchronizer and desynchronizer increases positive and negative correlation respectively between two SNs.
Given two input SNs $X$ and $Y$, the SC synchronizer (Fig.~\ref{fig:synchronizer}) produces two SNs $X'$ and $Y'$ which are more positively correlated and have the same value as the input SNs $X$ and $Y$ respectively.
The key idea is to dynamically pair up 1s and 0s from the two input streams as often as possible between the two input SNs.
When $X$ and $Y$ inputs are the same, the synchronizer simply passes them to the outputs.
If $X$ and $Y$ are different, the synchronizer ``saves'' the unpaired bit to be paired with a complement at a later SN index.
Pairing up bits as often as possible effectively induces positive correlation between the two SNs.

Our desynchronizer relies on a similar principle to increase negative correlation between two input SNs (Fig.~\ref{fig:desynchronizer}).
Instead of trying to dynamically pair up bits, the desynchronizer deliberately tries to \textit{unpair} bits and emit as many unpaired bits as possible.
To do this, when the $X$ and $Y$ inputs are both 1 the desynchronizer saves one of the bits and passes the other.
If the inputs are both zero, it takes a saved bit if available and overrides one of the zeros.
Finally, when input bits are different, the synchronizer simply passes them since they are already unpaired.
By unpairing as many bits as possible, the resulting SNs become negatively correlated.

To quantify their efficacy, we measure average SCC before and after each design, and the average \textit{bias}.
Bias is defined as the deviation in the output values from the input values.
Ideally the bias should be zero since our circuits should only alter the SN correlation and not SN value.
Table~\ref{tab:correlation} shows the initial and resulting SCCs, and average bias of result SNs for several RNG configurations averaged over all possible input values.
To measure induced correlation, we deliberately chose RNGs that initialize input SNs to be minimally correlated for the synchronizer and desynchronizer.
We also provide one RNG configuration where the inputs are initially positively correlated to test the desynchronizer.
In general, we find that our synchronizer and desynchronizer designs work well across all configurations.
We also observe stronger induced correlation when the input SNs have low discrepancy (i.e. generated with Halton and VDC RNGs).
The results also show that on average there is a slight negative bias in the resulting SNs.
This is due to saved bits getting ``stuck'' in the FSM at the end of the computation.

\begin{table}[t]
  \centering
  \caption{Average SN correlation before and after correlation manipulating circuits (N = 256).} 
  \label{tab:correlation}
  \begin{tabular}{@{}c|cc|cc|cc@{}} \toprule
    Design &
    \begin{tabular}{@{}c@{}}X\\RNG\end{tabular} &
      \begin{tabular}{@{}c@{}}Y\\RNG\end{tabular} &
    \begin{tabular}{@{}c@{}}Input\\SCC\end{tabular} &
      \begin{tabular}{@{}c@{}}Output\\SCC\end{tabular} &
        \begin{tabular}{@{}c@{}}X'\\Bias\end{tabular} &
          \begin{tabular}{@{}c@{}}Y'\\Bias.\end{tabular} \\ \midrule

          \multirow{3}{*}{\begin{tabular}{@{}c@{}}Synchronizer\\(Fig.~\ref{fig:synchronizer})\end{tabular}} & VDC & Halton
            & -0.048 & 0.996 & -0.001 & -0.002 \\
             & LFSR & VDC
              & -0.062 & 0.903 & -0.002 & -0.001 \\
               & Halton  & Halton
              & 0.984 & 0.992 & -0.002 & -0.002 \\ \hline
              \multirow{3}{*}{\begin{tabular}{@{}c@{}}Desynchronizer\\(Fig.~\ref{fig:desynchronizer})\end{tabular}} & VDC & Halton
                & -0.048 & -0.981 & -0.002 & 0 \\

                 & LFSR & VDC
                  & -0.062 & -0.788 & -0.002 & 0 \\
                   & Halton & Halton
                  & 0.984 & -0.930 & -0.003 & 0 \\
                  \hline

                  \multirow{3}{*}{\begin{tabular}{@{}c@{}}Decorrelator\\(Fig.~\ref{fig:shuffle-buffer}a)\end{tabular}}                    & LFSR & LFSR & 0.992 & 0.249 & 0.000 & -0.004 \\
                  & VDC & VDC & 0.992 & 0.168 & 0.001 & 0.003 \\
                   & Halton & Halton & 0.984 & 0.067 & 0.001 & 0.002 \\
                  \hline
                  \multirow{3}{*}{\begin{tabular}{@{}c@{}}Isolator\\Insertion\end{tabular}}     & LFSR   & LFSR   & 0.992 & 0.600  & -0.002 & 0.000 \\
                       & VDC    & VDC    & 0.992 & -0.637 & -0.004 & 0.000 \\
                       & Halton & Halton & 0.984 & -0.353 & 0.002 & 0.000 \\ \hline
                  \multirow{3}{*}{\begin{tabular}{@{}c@{}}Tracking\\Forecast\\Memory~\cite{tracking-forecast-memories}\end{tabular}}     & LFSR   & LFSR   & 0.992 & 0.654  & -0.014 & -0.051 \\
                       & VDC    & VDC    & 0.992 & 0.779 & 0.246 & 0.363 \\
                       & Halton & Halton & 0.984 & 0.353 & -0.005 & -0.007 \\
                  \bottomrule
  \end{tabular}
\end{table}

\subsection{Generalized Designs and Composition}

\noindent The synchronizer and desynchronizer designs can both be generalized or composed to improve the strength of the induced correlation.
To generalize these designs, the key idea is to extend the FSMs to save additional bits.
For the synchronizer, this is equivalent to adding an equal number of states to the left and right of the FSM to track how many bits from $X$ and $Y$ have been saved.
For the desynchronizer, we add additional FSM states and transitions to the FSM cycle to represent other possible saved bit configurations.
We refer to the number of bits that a synchronizer and desynchronizer can save as the save depth $D$.
The key idea is that by having a larger save depth, the designs will be more resilient to runs of 1s and 0s which reduce their efficacy.

These extensions present their own challenges: for designs with large $D$, it is possible for the saved bits to get ``stuck'' in the FSM before they can all be paired or unpaired.
In adversarial cases, this can result in larger biases from the original value if the saved bits in the FSM are not emitted before the end of the SN.
To mitigate this issue, one could add an optional FSM flush functionality which requires keeping track of the current offset into the bitstream $t$.
If the number of remaining saved bits in the FSM is greater than or equal to $t$, it forces the FSM to emit the saved bits regardless of saved state.
However, these modifications require additional overheads and can become tremendously expensive for large save depth $D$.

An alternative way to enhance correlation strength is to compose multiple synchronizers or desynchronizers with minimal depth $D = 1$ together in series.
Each synchronizer or desynchronizer will improve the correlation albeit with diminishing returns.
In the limit, output SNs will eventually become maximally correlated.
However, like the FSM extensions, it is still possible for residual bits to remain in each FSM resulting in compounding bias errors.
To address this issue, the FSMs can be modified to start with a saved $X$ or $Y$ bit by adjusting the initial state.

\subsection{Decorrelator}

\noindent We now introduce our decorrelator which is designed to take two SNs and emit two less correlated SNs.
Our decorrelator is composed of two shuffle buffers (Fig.~\ref{fig:shuffle-buffer}b).
A shuffle buffer is a small memory which randomly replaces and emits bits.
At each cycle, the shuffle buffer will either pass the current input, or store the current input and emit a previously stored input.
The depth $D$ of the memory can be parametrized; intuitively, a deeper memory allows the shuffle buffer to scramble bits across longer segments of the SN.
To determine which bit to store and emit at a given cycle, the decorrelator requires an auxiliary RNG input.
To decorrelate two different SNs, we use two shuffle buffers with different RNGs.

{
  \setlength{\belowcaptionskip}{-10pt}
\begin{figure}[b]
  \centering
  \includegraphics[width=\linewidth]{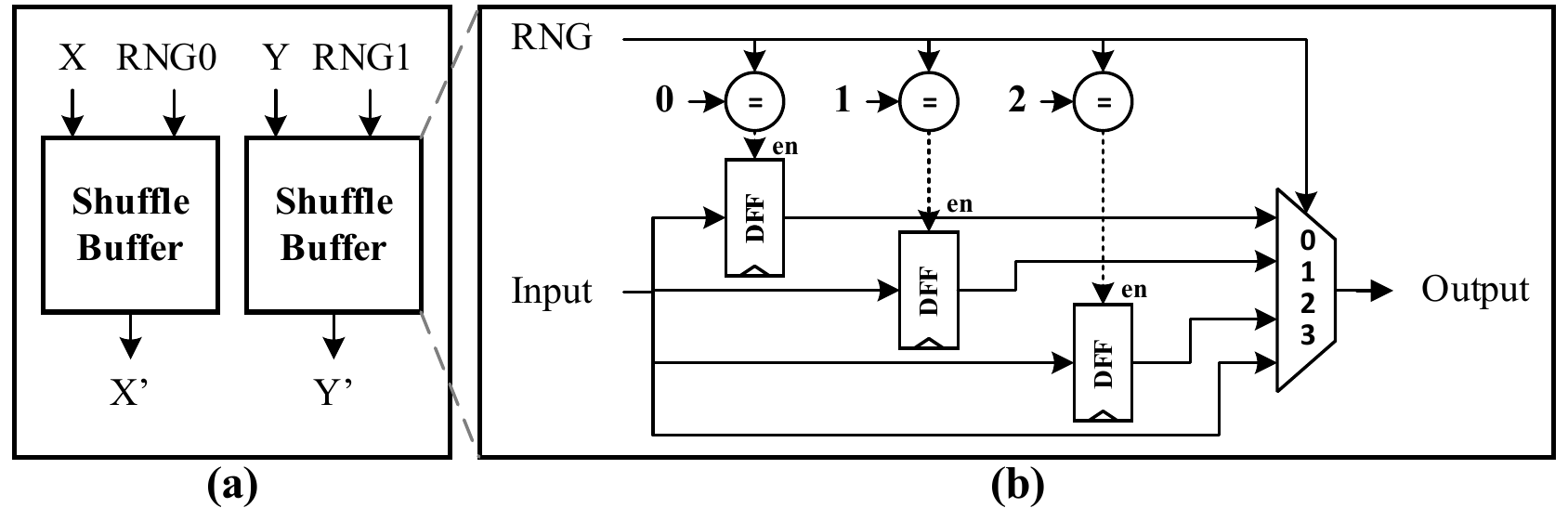}
  \caption{(a) Decorrelator design with (b) shuffle buffer depth $D = 4$.}
  \label{fig:shuffle-buffer}
\end{figure}
}
The efficacy of the decorrelator design for two SNs is shown in Table~\ref{tab:correlation} for several correlated SN configurations.
On average, we find that the decorrelator significantly reduces the correlation between SN operands.
To reduce bias errors, we initialize half of the buffer in the design to 1s and the others to 0s so that on average fewer 1s from the input SNs will get ``stuck'' in the buffer.
Like the synchronizer and desynchronizer, decorrelator designs can be composed in series to further reduce correlation.
We also evaluate isolators and tracking forecast memories~\cite{tracking-forecast-memories} for decorrelating SNs, and find they are both less effective than our proposed decorrelator.


\subsection{Improving SC Operations}

\begin{figure}[b]
  \centering
  \includegraphics[width=\linewidth]{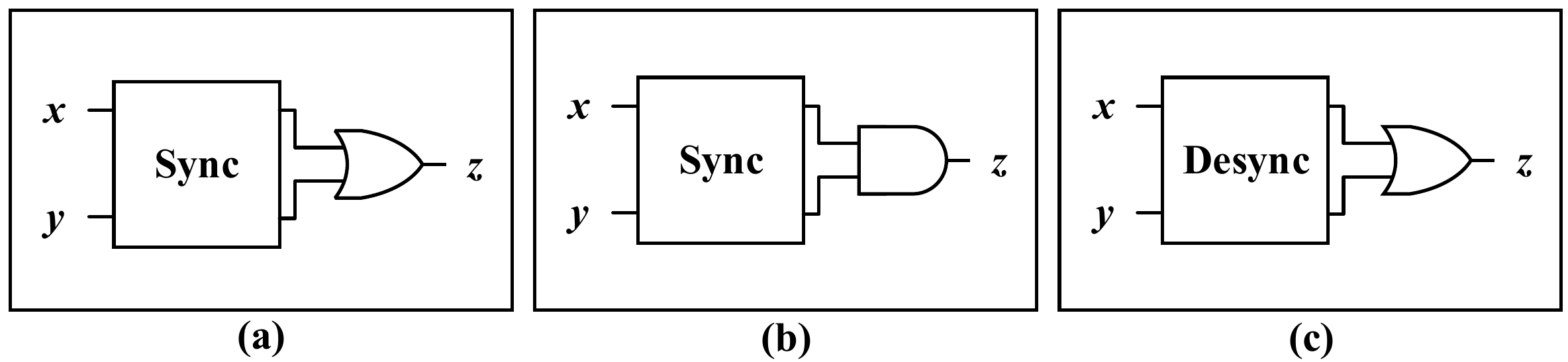}
  \caption{Improved SC circuits: (a) maximum, (b) minimum, and (c) saturating add.}
  \label{fig:new_designs}
\end{figure}

\noindent We now show how we can improve SC operations using our new correlation manipulating circuits for SC maximum, minimum, and saturating add which were originally proposed in~\cite{sc-correlation}.
Our improved SC maximum combines the synchronizer design with an OR gate (Fig.~\ref{fig:new_designs}a) while the SC minimum combines the synchronizers with an AND gate (Fig.~\ref{fig:new_designs}b).
For the SC maximum, the key insight is that for input SNs with maximal, positive correlation, the larger SN will exactly mask the smaller SN when taking the OR of the two SNs; the OR gate will also propagate any extra 1s in the larger SN.
If input SNs are not positively correlated, the resulting SN will be less accurate and have a value strictly \textit{greater than or equal to} the larger input SN.
Thus, we use the new synchronizer design to induce positive correlation between the two input SNs before feeding the them to the OR gate.
A similar idea is used to calculate the minimum between two SNs.
Instead of an OR gate, we use an AND gate to pass \textit{at most} the maximum number of masked 1s between the SNs; this results in a SN with the minimum value.
Finally, we prepend a desynchronizer to the OR gate to improve the SC saturating adder which requires negatively correlated inputs (Fig.~\ref{fig:new_designs}c).

To evaluate accuracy, we exhaustively generate all possible inputs using two uncorrelated RNGs - a VDC sequence and Halton sequence with base 3 - and report average absolute error.
We also evaluate the design power, area, energy, and accuracy across several maximum designs: (1) single OR gate, (2) our improved synchronizer-based maximum, and (3) the correlation agnostic maximum (CA maximum) in~\cite{sc-dcnn}.
For SC minimum, we compare (1) a single AND gate and (2) our synchronizer-based minimum.

Table~\ref{tab:ops} compares our improved synchronizer-based maximum and minimum against prior work, and shows that our designs are more accurate than simply using an OR gate without correlation manipulation.
Compared to the CA maximum, our synchronizer-based design is 5.2$\times$ smaller and 11.6$\times$ more energy efficient with minimal accuracy loss.
Using a deeper save depth can improve accuracy but also increases power, area, and energy relative to other designs.
As with many aspects of SC, this illustrates a design trade off: more accurate SC functional units are larger and consume more power.
We also observe similar results for the synchronizer-based minimum design.

\begin{table}[t]
  \caption{Average absolute error, bias, area, power, and energy for SC maximum and minimum designs for N = 256.}
  \label{tab:ops}
 \centering
  \begin{tabular}{@{}cccccc@{}} \toprule
    Design & \begin{tabular}{@{}c@{}}Abs. Error\end{tabular} &
      Avg. Bias &
        \begin{tabular}{@{}c@{}} Area  \\(\textmu m$^2$/op)\end{tabular} &
        \begin{tabular}{@{}c@{}} Power \\(\textmu W/op)\end{tabular} &
        \begin{tabular}{@{}c@{}} Energy\\(pJ/op)\end{tabular} \\ \midrule
        OR Max. & 0.087 & 0.087   & 2.16 & 0.26 & 165 \\
        CA Max. & 0.006 & 0.001   & 252.36 & 56.7 & 36288 \\
        \begin{tabular}{@{}c@{}}Sync. Max.\end{tabular} &
                  0.003 & 0.003   & 48.6 & 4.89 & 3130 \\
        AND Min. & 0.082 & -0.082 & 2.16 & 0.25 & 158 \\
        Sync. Min. & 0.005 & 0.005 & 45.0 & 8.38 & 5363 \\
        \bottomrule
  \end{tabular}
\end{table}

\section{Image Processing Case Study} \label{sec:evaluation}

\noindent This section evaluates the relative power, area, energy, and accuracy of an image processing pipeline to illustrate the merits of our new correlation manipulating circuits.

\subsection{Methodology}

\noindent We evaluate the power, area, accuracy, and energy of a Gaussian blur (GB) followed by a Roberts cross edge detector (ED)~\cite{sc-image-processing} SC accelerator for several configurations.
This image processing pipeline illustrates the role of correlation manipulation since the SC Gaussian blur requires inputs to be uncorrelated while the edge detector requires positively correlated inputs.
Our accelerator architecture expects the input image to be tiled and processes each tile individually one at a time.
All outputs within each tile are computed in parallel before moving on to the next tile.
For this evaluation, we use a 10$\times$10 input tile size.

We evaluate several SC accelerator variants: (1) an accelerator with no correlation manipulating circuits between GB and ED, (2) an accelerator that uses regeneration between the GB and ED, and (3) an accelerator that uses synchronizers to manipulate correlation between GB and ED.
For each design, we synthesize, place, and route each accelerator using a TSMC 65nm library with Synopsys Design Compiler, IC Compiler, and PrimeTime.
For power measurements, we use post-placement and route simulation using random traces to make measurements data agnostic.
To model quality, we use a cycle-level simulator which uses models that have been verified against RTL simulation traces.
We report accuracy in terms of the average absolute error of the SC result compared to a floating point baseline image.

\subsection{Evaluation Results}

{
\begin{table}[t]
  \centering
  \caption{Image results for GB followed by ED. Using synchronizers is more energy efficient than using regeneration.}
  \label{tab:gb_ed}
  \begin{tabular}{@{}m{0.10\linewidth}m{0.17\linewidth}m{0.18\linewidth}m{0.18\linewidth}m{0.18\linewidth}@{}} \toprule
    \centering Design &
    \centering Floating Point &
    \centering SC No Manipulation &
    \centering SC Regeneration &
    \centering SC Synchronizer \tabularnewline \midrule

    \centering
      \begin{tabular}{@{}c@{}} Image\\Result \end{tabular}
      &
    \centering \begin{tabular}{@{}c@{}}\includegraphics[width=1.5cm]{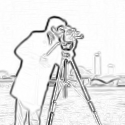} \end{tabular}&
    \includegraphics[width=1.5cm]{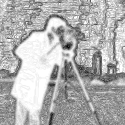} &
    \includegraphics[width=1.5cm]{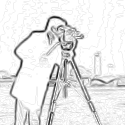} &
    \includegraphics[width=1.5cm]{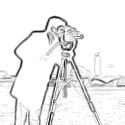} \tabularnewline

    \hline



        \centering \begin{tabular}{@{}c@{}}Area\end{tabular} &
          \centering -- &
          \centering 24313 \textmu m$^2$&
          \centering 34802 \textmu m$^2$&
          \centering 36202 \textmu m$^2$
          \tabularnewline

          \hline

          \centering \begin{tabular}{@{}c@{}}Energy\end{tabular} &
            \centering -- &
            \centering 1383 nJ/frame &
            \centering 1971 nJ/frame &
            \centering 1505 nJ/frame
            \tabularnewline

            \hline

            \centering \begin{tabular}{@{}c@{}} Abs.\\Error\end{tabular} &
              \centering 0 &
              \centering 0.076 &
              \centering 0.019 &
              \centering 0.020
              \tabularnewline

            \bottomrule
  \end{tabular}
\end{table}
}

\noindent Table~\ref{tab:gb_ed} shows the quality, energy, and area results for the combined GB and ED accelerator.
We find the image quality and average absolute error is markedly better when using regeneration or synchronizers between image processing kernels than the results generated by the design without correlation manipulating circuits.
In terms of energy, our synchronizer-based design improves total accelerator energy consumption by 24\% over the design using regeneration.
We also find the average absolute error difference between the design using regeneration versus our new design is negligible.

We also compare the overheads of synchronizer circuits and regeneration.
To do this, we tabulate the power break down for S/D converters, D/S converters, compute kernels, RNGs, and synchronizers.
We then aggregate the costs associated only with correlation manipulation.
The two designs require different numbers of S/D and S/D converters, and synchronizers to accomplish the same task; in this case, our synchronizer-based design requires 2$\times$ more synchronizers than the number of S/D and D/S converters used by regeneration.
This is because synchronizers only induce correlation between two SNs while using regeneration induces correlation between \textit{all} SNs.
Fortunately, for many applications like the ED kernel, it is sufficient to induce correlation between pairs of SNs.
As a result, we find that the overhead of correlation manipulation using synchronizers is $3.0\times$ more energy efficient than when using regeneration.

\section{Related Work}\label{sec:related-work}

\noindent We are not the first to exploit correlation manipulating circuits to improve the fidelity of SC computation.
Ting and Hayes~\cite{ting16} introduce an algorithm for placing isolator circuits to decorrelate SNs.
However, isolators are limited in that they only shift bits by a fixed offset and do not scramble relative bit order; our decorrelator can scramble bit order across larger segments of the SN.
Tehrani et al.~\cite{edge-memories, tracking-forecast-memories} propose edge memories and tracking forecast memories (TFM) for improving SN correlation in low-density parity-check (LDPC) decoding.
Since TFMs were designed specifically for LDPC decoding, they do not generalize as well as our new decorrelator design; TFMs are also larger since portions of the TFM require binary-encoded arithmetic.
Finally, Parhi et al.~\cite{parhi15} propose a technique for synthesizing correlated and uncorrelated SNs but do not propose generic correlation manipulating circuits.

\section{Conclusions}
\label{sec:conclusions}

\noindent This paper presents a new set of correlation manipulating circuits - a synchronizer, desynchronizer, and decorrelator - for managing correlation between SNs.
We propose improved SC maximum, minimum, and saturating adder designs based on our correlation manipulating circuits, and show that they are smaller, more accurate, and lower power than previous designs.
Finally, we show how using correlation manipulating circuits can improve the accuracy of SC computations and are more energy efficient than using existing correlation manipulation techniques in the context of an image processing pipeline.

\section{Acknowledgements}

\noindent This work was supported in part by the National Science Foundation Grant CCF-1518703, generous gifts from Oracle Labs, and by C-FAR, one of the six SRC STARnet Centers, sponsored by MARCO and DARPA.

\widowpenalty10000


\balance
\bibliographystyle{IEEEtran} 
\bibliography{IEEEabrv,references}

\begin{thebibliography}{10}
\providecommand{\url}[1]{#1}
\csname url@samestyle\endcsname
\providecommand{\newblock}{\relax}
\providecommand{\bibinfo}[2]{#2}
\providecommand{\BIBentrySTDinterwordspacing}{\spaceskip=0pt\relax}
\providecommand{\BIBentryALTinterwordstretchfactor}{4}
\providecommand{\BIBentryALTinterwordspacing}{\spaceskip=\fontdimen2\font plus
\BIBentryALTinterwordstretchfactor\fontdimen3\font minus
  \fontdimen4\font\relax}
\providecommand{\BIBforeignlanguage}[2]{{%
\expandafter\ifx\csname l@#1\endcsname\relax
\typeout{** WARNING: IEEEtran.bst: No hyphenation pattern has been}%
\typeout{** loaded for the language `#1'. Using the pattern for}%
\typeout{** the default language instead.}%
\else
\language=\csname l@#1\endcsname
\fi
#2}}
\providecommand{\BIBdecl}{\relax}
\BIBdecl

\bibitem{gaines69}
B.~R. Gaines, \emph{Stochastic Computing Systems}.\hskip 1em plus 0.5em minus
  0.4em\relax Boston, MA: Springer US, 1969, pp. 37--172.

\bibitem{sc-correlation}
A.~Alaghi and J.~P. Hayes, ``Exploiting correlation in stochastic circuit
  design,'' in \emph{2013 IEEE 31st International Conference on Computer Design
  (ICCD)}, Oct 2013, pp. 39--46.

\bibitem{sc-apc}
P.~S. Ting and J.~P. Hayes, ``Stochastic logic realization of matrix
  operations,'' in \emph{2014 17th Euromicro Conference on Digital System
  Design}, Aug 2014, pp. 356--364.

\bibitem{gaines67}
B.~R. Gaines, ``Stochastic computing,'' in \emph{Proceedings of the April
  18-20, 1967, Spring Joint Computer Conference}, ser. AFIPS '67
  (Spring).\hskip 1em plus 0.5em minus 0.4em\relax New York, NY, USA: ACM,
  1967, pp. 149--156.

\bibitem{alaghi14-date}
A.~Alaghi and J.~P. Hayes, ``Fast and accurate computation using stochastic
  circuits,'' in \emph{2014 Design, Automation Test in Europe Conference
  Exhibition (DATE)}, March 2014, pp. 1--4.

\bibitem{sc-divider}
T.~H. Chen and J.~P. Hayes, ``Design of division circuits for stochastic
  computing,'' in \emph{2016 IEEE Computer Society Annual Symposium on VLSI
  (ISVLSI)}, July 2016, pp. 116--121.

\bibitem{low-discrepancy-sequences}
A.~Alaghi and J.~P. Hayes, ``Fast and accurate computation using stochastic
  circuits,'' in \emph{2014 Design, Automation Test in Europe Conference
  Exhibition (DATE)}, March 2014, pp. 1--4.

\bibitem{liu17-date}
S.~Liu and J.~Han, ``Energy efficient stochastic computing with sobol
  sequences,'' in \emph{Design, Automation Test in Europe Conference Exhibition
  (DATE), 2017}, March 2017, pp. 650--653.

\bibitem{lee17-date}
V.~T. Lee, A.~Alaghi \emph{et~al.}, ``Energy-efficient hybrid stochastic-binary
  neural networks for near-sensor computing,'' in \emph{Design, Automation Test
  in Europe Conference Exhibition (DATE), 2017}, March 2017, pp. 13--18.

\bibitem{ting16}
P.~S. Ting and J.~P. Hayes, ``Isolation-based decorrelation of stochastic
  circuits,'' in \emph{2016 IEEE 34th International Conference on Computer
  Design (ICCD)}, Oct 2016, pp. 88--95.

\bibitem{tracking-forecast-memories}
S.~S. Tehrani, A.~Naderi \emph{et~al.}, ``Tracking forecast memories in
  stochastic decoders,'' in \emph{2009 IEEE International Conference on
  Acoustics, Speech and Signal Processing}, April 2009, pp. 561--564.

\bibitem{sc-dcnn}
A.~Ren, Z.~Li \emph{et~al.}, ``Sc-dcnn: Highly-scalable deep convolutional
  neural network using stochastic computing,'' in \emph{Proceedings of the
  Twenty-Second International Conference on Architectural Support for
  Programming Languages and Operating Systems}, ser. ASPLOS '17.\hskip 1em plus
  0.5em minus 0.4em\relax New York, NY, USA: ACM, 2017, pp. 405--418.

\bibitem{sc-image-processing}
A.~Alaghi, C.~Li \emph{et~al.}, ``Stochastic circuits for real-time
  image-processing applications,'' in \emph{Proceedings of the 50th Annual
  Design Automation Conference}, ser. DAC '13.\hskip 1em plus 0.5em minus
  0.4em\relax New York, NY, USA: ACM, 2013, pp. 136:1--136:6.

\bibitem{edge-memories}
S.~S. Tehrani, W.~J. Gross \emph{et~al.}, ``Stochastic decoding of ldpc
  codes,'' \emph{IEEE Communications Letters}, vol.~10, no.~10, pp. 716--718,
  Oct 2006.

\bibitem{parhi15}
M.~Parhi, M.~D. Riedel \emph{et~al.}, ``Effect of bit-level correlation in
  stochastic computing,'' in \emph{2015 IEEE International Conference on
  Digital Signal Processing (DSP)}, July 2015, pp. 463--467.

\end{thebibliography}

\end{document}